\documentclass[preprint,authoryear,11pt,3p]{elsarticle}

\usepackage[T1]{fontenc}
\usepackage{setspace}
\usepackage{amssymb,multirow,amsmath,graphicx,graphics,array,lscape,rotating,color,amsthm,ifsym,courier,listings,enumitem,hyperref,wrapfig,bm,tikz-qtree,tikz,bbding,arydshln,bbm}

\definecolor{dkgreen}{rgb}{0,0.6,0}
\definecolor{gray}{rgb}{0.5,0.5,0.5}
\definecolor{mauve}{rgb}{0.58,0,0.82}
\definecolor{red}{rgb}{0.8,0,0}
\definecolor{lblue}{rgb}{0.2,0.45,0.6}
\lstdefinestyle{mystyle}{frame=tb,
  language=R,
  aboveskip=3mm,
  belowskip=3mm,
  showstringspaces=false,
  columns=flexible,
  basicstyle={\footnotesize\ttfamily},
  numbers=none,
  numberstyle=\tiny\color{gray},
  keywordstyle=\color{blue},
  commentstyle=\color{dkgreen},
  stringstyle=\color{mauve},
  breaklines=true,
  breakatwhitespace=true,
  tabsize=3
}
\usetikzlibrary{arrows,shapes,positioning,shadows,trees}
\tikzset{
  basic/.style  = {draw, text width=5cm, drop shadow, font=\sffamily, rectangle},
  root/.style   = {basic, rounded corners=6pt, thin, align=center, fill=white!30, text width=6em, sibling distance=60mm},
   level 1/.style = {basic, rounded corners=6pt, thin,align=center, fill=white!60, text width=6em, sibling distance=70mm},
  level 2/.style = {basic, rounded corners=6pt, thin,align=center, fill=white!60, text width=6em, sibling distance=25mm},
  level 3/.style = {basic, rounded corners=6pt, thin,align=center, fill=white!60, text width=4em, sibling distance=10mm},
  level 4/.style = {basic, rounded corners=6pt, thin,align=center, fill=white!60, text width=6em, sibling distance=40mm},
  level 5/.style = {basic, rounded corners=6pt, thin,align=center, fill=white!60, text width=5em, sibling distance=30mm},
}

\journal{European Journal of Operational Research}

\begin{document}

\begin{frontmatter}

\title{Reconciliation of probabilistic forecasts \\
with an application to wind power}

\author[label1,label2]{Jooyoung Jeon\corref{cor1}}
\ead{j.jeon@bath.ac.uk}
\cortext[cor1]{Correspondence: Jooyoung Jeon}

\author[label3]{Anastasios Panagiotelis}
 \ead{Anastasios.Panagiotelis@monash.edu}

\author[label1]{Fotios Petropoulos}
\ead{f.petropoulos@bath.ac.uk}

\address[label1]{School of Management, University of Bath}
\address[label2]{Graduate School of Engineering Practice, Seoul National University}
\address[label3]{Department of Econometrics \& Business Statistics, Monash University}

\begin{abstract}
New methods are proposed for adjusting probabilistic forecasts to ensure coherence with the aggregation constraints inherent in temporal hierarchies. The different approaches nested within this framework include methods that exploit information at all levels of the hierarchy as well as a novel method based on cross-validation. The methods are evaluated using real data from two wind farms in Crete, an application where it is imperative for optimal decisions related to grid operations and bidding strategies to be based on coherent probabilistic forecasts of wind power. Empirical evidence is also presented showing that probabilistic forecast reconciliation improves the accuracy of both point forecasts and probabilistic forecasts.
\end{abstract}

\end{frontmatter}
 
\section{Introduction}\label{sec:intro}
Data are often arranged in cross-sectional or temporal hierarchies characterised by an aggregation structure that holds for all realised values; for example, the annual sum of monthly data series will be equivalent to annual data series. When forecasts are independently produced for different series or levels within a hierarchy these aggregation constraints will not hold, a property known as {\em incoherence}.  To ensure that operational decisions are aligned, a rich literature has emerged on forecast reconciliation \citep{Athanasopoulos2009-db,Hyndman2011OptimalSeries,Athanasopoulos2017ForecastingHierarchies, Wickramasuriya2017}.  These methods not only ensure that forecasts are coherent but also lead to improvements in forecast accuracy. However, a shortcoming of these methods is their focus on point forecasting despite the increasing importance of probabilistic forecasts on decision-making \citep{Gneiting2014ProbabilisticForecasting}.  This paper proposes a new methodology for the reconciliation of probabilistic forecasts.

Our proposed methodology can be described according to its three novel features.  First, this study is the first to combine information about the full probabilistic forecast of each series in the reconciliation process.  Second, this study is the first to focus on producing coherent probabilistic forecasts in the temporal rather than in the cross-sectional hierarchical setting, although we note that our methodology is general enough to handle both of these settings.  Third, this study is the first to consider training reconciliation weights via a cross-validation procedure in either the point or probabilistic forecasting setting.  Indeed to the best of our knowledge, the only other paper to tackle the issue of coherent probabilistic forecasts is that of \citet{Taieb2017CoherentSeries} and our approach can be distinguished from theirs by each of the above-mentioned features.  Crucially, with the exception of the mean and variance, the construction of a coherent probabilistic forecast by \citet{Taieb2017CoherentSeries} relies on a bottom up approach. In contrast our entire reconciled probabilistic forecast is based on probabilistic forecasts of series from all hierarchical levels.

The methods we propose are evaluated using wind power data measured at various frequencies ranging from hourly to daily. This application is chosen for two main reasons.  First, due to the highly volatile nature of wind power generation, informed decision-making depends not only on point forecasts but on probabilistic considerations. For instance, dispatch and risk management decisions may be based on the probability that a wind farm supplies at least 300{\em kWh} between midnight and 6am the following day.  Second, wind farm operators, grid system operators and electricity traders are each required to make decisions based on different forecast horizons and sampling intervals. As such coherent probabilistic forecasts are crucial to ensure aligned decision-making.  Our empirical results demonstrate that the proposed reconciliation methods improve the accuracy of point and probabilistic forecasts, with more substantial improvements at higher aggregation levels. 

In the next section, we review the literature on hierarchical forecast reconciliation. Section \ref{sec:newmethod} presents the methods to produce coherent and reconciled density forecasts. Section \ref{sec:design} introduces our wind power data, and describes the density forecasting models for wind power generation. Section \ref{sec:results} describes the empirical results of the various reconciliation methods considered in Section \ref{sec:newmethod}. The final section provides a summary and conclusion.

\section{Background}\label{sec:literature}

\subsection{Cross-sectional Hierarchical Reconciliation of Point Forecasts}

\begin{figure}
 \begin{center}
        \begin{tikzpicture}
        \node[root] {Company}        
          child {node[level 2] (c1) {Category 1}
          child {node[level 3] (c11) {SKU 1}}
          child {node[level 3] (c12) {SKU 2}}
          child {node[level 3] (c13) {SKU 3}}
          }
          child {node[level 2] (c2) {Category 2}                       
          child {node[level 3] (c21) {SKU 4}}
          child {node[level 3] (c22) {SKU 5}}
          };
        \end{tikzpicture}
    \caption{A cross-sectional hierarchy.}  
\label{fig:chierarchy}
 \end{center}
\end{figure}
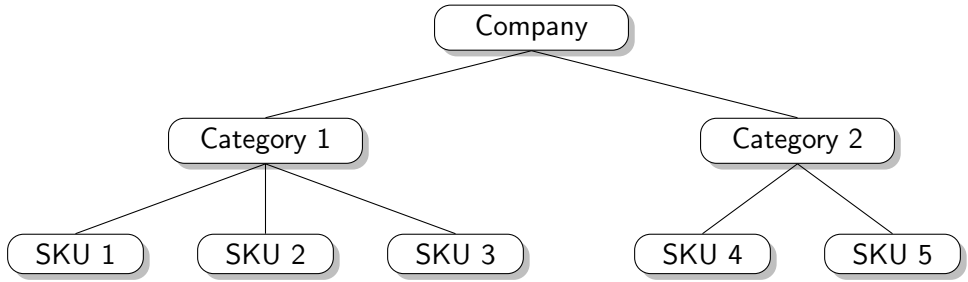

Data within companies are organised in hierarchical structures. For example, a company may organise its five stock keeping units (SKUs) into two categories, as depicted in Figure \ref{fig:chierarchy}. If the historical data at the bottom level (SKU) are available, then data at every other level can be calculated using appropriate aggregations. Forecasts may be produced at any of the three levels of the hierarchy. However, if forecasts are independently produced at all levels they will not be coherent. For example, the sum of the forecasts of SKUs 1, 2 and 3 in Figure \ref{fig:chierarchy} is not guaranteed to be the same as the forecast of Category 1. 

One way to tackle this issue is to simply produce forecasts on a single hierarchical level. For example, forecasts can be produced only on the very bottom level, and then aggregated to the higher levels in the hierarchical structure, an approach known as the \textit{bottom-up} approach \citep[see for example][]{Dangerfield1992-hd,Zellner2000-vi,Athanasopoulos2009-db}. In some cases, the bottom-level data may be too granular or noisy, rendering the forecasting task difficult. Alternatively, forecasts may be produced at the very top-level and then appropriately disaggregated to lower level forecasts, an approach known as \textit{top-down} \citep{Lutkepohl1984-qk,Fliedner1999-fx,Gross1990-bl}. Disaggregation of the forecasts to lower levels may be based on historical or predicted proportions of the lower level data \citep{Athanasopoulos2009-db}. The top-down approach has the disadvantage of information loss, as aggregated series may not reflect the individual characteristics of their descendants. Finally, forecasts can also be produced at a middle level; forecasts for higher/lower levels nodes can be calculated by appropriate aggregation/disaggregation of the middle-level forecasts. This approach is known as \textit{middle-out}, a conceptual combination of the bottom-up and top-down approaches.

A shortcoming of the methods above is that forecasts are only based on information at a single level of the hierarchy.  The optimal combination method introduced by \cite{Athanasopoulos2009-db} and \cite{Hyndman2011OptimalSeries}  overcome this problem  by tackling hierarchical forecasting in two stages.  In the first stage, forecasts are produced for all series at all levels independently.  In the second stage, these forecasts are adjusted in a reconciliation step to ensure coherence with aggregation constraints.  More specifically the reconciled forecast for each node is formed as a weighted combination of the original - or so-called `base' - forecasts of all nodes, in a way that ensures coherence for the hierarchy overall.  The key advantage of reconciliation is that information is used at all levels of the hierarchy in contrast to the approaches described in the previous paragraph that focus on a single level. More recently, \cite{Hyndman2016-xw} propose algorithms for fast computation of coherent hierarchical forecasts, and \cite{Wickramasuriya2017} suggest calculating coherent forecasts through trace minimisation.

\subsection{Temporal Hierarchical Reconciliation of Point Forecasts}
A time series can be aggregated or disaggregated to create alternative frequency (or resolution) as needed. Time series at different frequencies will exhibit different characteristics. Seasonality and noise will be amplified in lower aggregation levels (higher frequencies), while the long-term trend can be more easily estimated using higher aggregation levels (lower frequencies) \citep{Kourentzes2014-jq, Spithourakis2012}. Similar to the case of cross-sectional aggregation, forecasts produced using data at different frequencies will not generally agree. For example, the sum of the forecasts for the next three months produced using data measured at the monthly frequency will not equal to the one-step-ahead quarterly forecast based on data measured at a quarterly frequency. This problem is particularly relevant for aligning decisions across the different departments within a company (operations, sales, finance, marketing, strategy), which usually operate at different data frequencies.

Similarly to cross-sectional aggregation, the issue of non-coherent forecasts at different temporal aggregation levels can be addressed either by combining (reconciling) the forecasts from multiple aggregation levels or by producing forecasts for a single temporal aggregation level and then deriving the forecasts at the other levels as discussed previously.

\cite{Nikolopoulos2011} show empirically that in the context of intermittent demand there exists an optimal aggregation level, unique to each series, and proposed the Aggregate-Disaggregate Intermittent Demand Approach (ADIDA), where forecasts are produced at a (single) higher aggregation level and the lower level forecast is subsequently produced by disaggregation. This approach is particularly relevant for slow moving data, as temporal aggregation will result in series with a lower degree of intermittence \citep{Petropoulos2016-uu}. \cite{Tabar2013} derive analytical results that improvement in forecasting performance is a function of the aggregation level, under specific data generation processes.

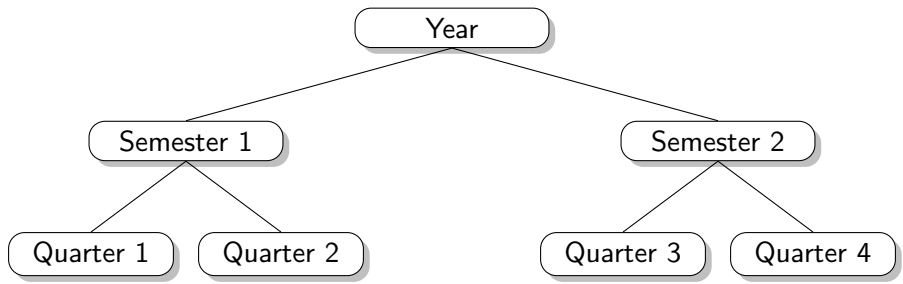
\begin{figure}
 \begin{center}
        \begin{tikzpicture}
        \node[root] {Year}        
          child {node[level 4] (c1) {Semester 1}
          child {node[level 5] (c11) {Quarter 1}}
          child {node[level 5] (c12) {Quarter 2}}
          }
          child {node[level 4] (c2) {Semester 2}                       
          child {node[level 5] (c21) {Quarter 3}}
          child {node[level 5] (c22) {Quarter 4}}
          };
        \end{tikzpicture}
    \caption{A temporal hierarchy.}  
\label{fig:thierarchy}
 \end{center}
\end{figure}

The idea of using aggregation/disaggregation for forecasting was further extended to derive the combined forecasts from forecasts simultaneously produced at multiple temporal aggregation (MTA) levels by \cite{Kourentzes2014-jq} and \cite{Petropoulos2014-ad}. MTA was also applied to the context of intermittent demand \citep{Petropoulos2015-vq}, and \cite{Kourentzes2016-bf} propose an extension of incorporating the effects of external variables. More recently, \cite{Athanasopoulos2017ForecastingHierarchies} express the MTA approach as a hierarchical concept using a temporal hierarchy for forecasting. A simple temporal hierarchy is depicted in Figure \ref{fig:thierarchy}, where the bottom-level data are at a quarterly frequency (1 quarter per node), middle-level data are at a semesterly frequency (2 quarters per node), and the top-level represents the yearly frequency (4 quarters for the top-level node). 

The representation of multiple temporal aggregation as temporal hierarchies allows for the application of the approaches designed for cross-sectional hierarchies, such as bottom-up, top-down, middle-out and optimal combination. Moreover, \cite{Athanasopoulos2017ForecastingHierarchies} provide three approximations of the sample covariance estimator of the covariance matrix of the base forecast errors. These approximations are based on hierarchy variance scaling, series variance scaling and structural scaling, an order that reflects on their increasing simplicity in terms of implementation. \cite{Athanasopoulos2017ForecastingHierarchies} show empirically that simpler scaling approximations provide better results, especially as the frequency of the bottom level increases. Note that in contrast to cross-sectional hierarchies where forecasts are produced separately for each node, forecasts within a temporal hierarchy are typically produced by fitting one model per aggregation level to model dependencies over time and the unique behaviour of each frequency. For example, a single model fitted to the quarterly time series produces multi-step ahead forecasts for quarters 1 to 4 at the lowest level in Figure \ref{fig:thierarchy}.

\subsection{Hierarchical Reconciliation of Probabilistic Forecasts}

Decision-making based on probabilistic forecasts has received increasing attention recently \citep[\ and references therein]{Gneiting2014ProbabilisticForecasting}. In a similar way to point forecasts, probabilistic forecasts could be produced independently for each level in the hierarchy, but independent series cannot be said to be coherent since the aggregation constraint induces dependence between the variables. The first approach to tackling hierarchical forecasting in the probabilistic setting is the paper of \cite{Taieb2017CoherentSeries}. After carrying out reconciliation on the mean, they construct a coherent probabilistic forecast in a bottom up fashion where the dependency between nodes at each level is modelled by reordering quantile forecasts as suggested by \cite{Arbenz2012CopulaReordering}. The method we propose is distinct from \cite{Taieb2017CoherentSeries} in two ways. First, our proposed method is a true reconciliation method, where each probabilistic forecast is based on information from all nodes in the hierarchy. Second, our problem focuses on temporal aggregation of density forecasts which provides a distinct case since dependence within each level can be obtained directly rather than through copula modelling. Recently, \cite{Athanasopoulos2017ForecastingHierarchies} propose methods to reconcile temporal point forecasts in the hierarchy, but none has yet focused on temporal hierarchical reconciliation of density forecasts. To the best of our knowledge, this study is the first to consider reconciliation of probabilistic forecasts for temporal hierarchies.

\section{Methodology}\label{sec:newmethod}

Let us introduce the following notation. We let $x^t_{j,\bm{f}_l}$ be the realisation of a variable recorded on cycle $t$ during the $j^{th}$ period of the cycle, where $\bm{f}_l$ is the sampling interval for level $l$ of the hierarchy. Cycle may refer, for instance, to a full year. For example, in the case of the hierarchy in Figure \ref{fig:thierarchy_notation}, we let $\bm{f} =[4,2,1]$. Subsequently, $x^1_{1,f_{1}}$ = $x^1_{1,4}$ is the demand for the first year (first four quarters), $x^3_{2,2}$ is the demand for the second semester of the third year and $x^5_{3,1}$ is the demand for the third quarter of the fifth year.  The same notation can be used for any other temporal hierarchy. Assuming, for example, a daily cycle and hourly data granularity, $\bm{f}=[24, 12, 8, 6, 4, 3, 2, 1]$ with $x^{10}_{5,4}$ referring to the 4 hourly demand of the fifth observation (16:00-20:00) of the tenth day. In the rest of Section \ref{sec:newmethod}, we will illustrate the methods of our interest using the temporal hierarchy depicted in Figure \ref{fig:thierarchy_notation}.


Let the scaled vectors $\bm{z}^t_{l}:=(\bm{f}_L/\bm{f}_l)(x^t_{1,\bm{f}_l},\ldots,x^t_{(\bm{f}_1/\bm{f}_l),\bm{f}_l})'$ for all $l$, where $L$ is the number of levels of the hierarchy ($L=3$ for the example hierarchy in Figure~\ref{fig:thierarchy_notation}). Then, $\bm{z}^t_{l}$ is the vector of the realisations of all the nodes at the level $l$, scaled to be in the same units as the lowest level $L$, i.e. the highest resolution. This allows us to avoid the complex scale conversion in the density reconciliation between any levels and to interpret reconciliation as forecast combination between levels. Afterwards, the probabilistic forecasts can be rescaled back to the original units for each level. Finally, let ${\bm y}^t:=({\bm {z}'^t_1},\ldots,{\bm {z}'^t_L})'$. The notation $y_i^{t}$ will be used to denote the $i^{th}$ scalar element of $\bm{y}^t$ for $i=1,\ldots,M$, where $M$ is the number of nodes in the hierarchy (e.g. $M=7$ in Figure \ref{fig:thierarchy_notation}).

\subsection{Coherent and Reconciled Probabilistic Forecasts}
\begin{figure}
 \begin{center}
        \begin{tikzpicture}
        \node[root] {$x^1_{1,4}$ or $y^1_1$}        
          child {node[level 4] (c1) {$x^1_{1,2}$ or $y^1_2$}
          child {node[level 5] (c11) {$x^1_{1,1}$ or $y^1_4$}}
          child {node[level 5] (c12) {$x^1_{2,1}$ or $y^1_5$}}
          }
          child {node[level 4] (c2) {$x^1_{2,2}$ or $y^1_3$}                   
          child {node[level 5] (c21) {$x^1_{3,1}$ or $y^1_6$}}
          child {node[level 5] (c22) {$x^1_{4,1}$ or $y^1_7$}}
          };
        \end{tikzpicture}
    \caption{An illustration of notation for a temporal hierarchy.}  
\label{fig:thierarchy_notation}
 \end{center}
\end{figure}
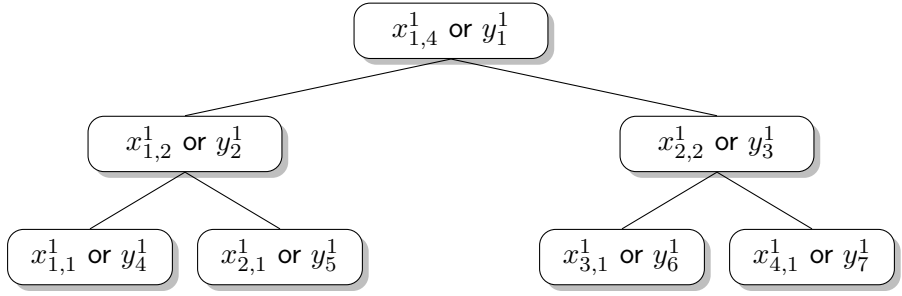

Some care must be taken in extending concepts such as coherent forecasts and reconciled forecasts to the probabilistic setting. Formal definitions of coherent probabilistic forecasts  are provided in \cite{Taieb2017CoherentSeries}. In brief, a coherent probabilistic forecast is an $M$-dimensional multivariate distribution which, due to the degeneracy induced by the aggregation constraints, is only supported on an $m$-dimensional linear subspace of $\mathbb{R}^M$ where $m=\bm{f}_1/\bm{f}_L<M$ (e.g. $m=4$ in Figure \ref{fig:thierarchy_notation}, simply referring to the number of nodes on the bottom level).

As discussed in Section~\ref{sec:literature}, reconciliation in the point forecasting context refers to a process by which a vector of incoherent forecasts is made coherent.  We now provide some detail.  Letting ${\hat{\bm y}}$ be a vector of unreconciled or `base' forecasts, then a reconciled point forecast is given by ${\tilde{\bm y}}={\bm S}{\bm P}{\hat{\bm y}}$.  The matrix ${\bm P}$ is a $m\times M$ matrix that forms point forecasts for the bottom level of the hierarchy as linear combinations of the base point forecasts of all nodes.  The matrix ${\bm S}$ is a $M\times m$ matrix that encodes the aggregation constraints and recovers a full set of coherent forecasts from the bottom level forecasts.  For the simple hierarchy in Figure~\ref{fig:thierarchy_notation}, ${\bm S}$ is given by
\begin{equation}
{\bm S}=\begin{bmatrix}
 1/4&1/4&1/4&1/4\\
 1/2&1/2&0&0\\
 0&0&1/2&1/2\\
 1&0&0&0\\
 0&1&0&0\\
 0&0&1&0\\
 0&0&0&1\\
 \end{bmatrix}
\end{equation}

\noindent Taken together, the matrix ${\bm S}{\bm P}$ is a projection matrix which takes any vector in $\mathbb{R}^M$ and projects it to an $m$-dimensional subspace spanned by the vectors of ${\bm S}$, a linear subspace where all aggregation constraints hold.

A common way to build probabilistic forecasts - that we follow here - is to generate a sample of size $N$ from the distribution $f\left({\bm {\bm y}^{t+h}}|\mathcal{F}^t_l;\hat{\bm{\theta}}\right)$, where $\mathcal{F}^t_l$ represents all the information up to time $t$ in the level $l$ and $\hat{\bm{\theta}}$ indicates that the probabilistic forecast is based on parameter estimates.  Denoting the $i^{th}$ vector from this sample as $\hat{\bm y}_i^{t+h|t}$, we can store these in a matrix as $\hat{\bm Y}=\left(\hat{\bm y}_1^{t+h|t},\ldots, \hat{\bm y}_N^{t+h|t}\right)$.  Typically there is no guarantee that the aggregation constraints will hold for each (or in fact any) of the columns of $\hat{\bm Y}$.  However, if $\hat{\bm Y}$ is pre-multiplied by a projection matrix to give $\tilde{\bm Y}={\bm S}{\bm P}\hat{\bm Y}$, the columns of the resulting matrix will respect the aggregation constraints and can therefore be thought of as observations sampled from the reconciled probabilistic forecast.  In this way existing reconciliation methods for the mean can be extended to a probabilistic setting.  To summarise, the process for forming probabilistic forecasts consists of two stages, in the first a sample is obtained from an estimate of the joint density $f\left({\bm {\bm y}^{t+h}}|\mathcal{F}^t_l;\hat{\bm{\theta}}\right)$, and in the second each sampled vector is pre-multiplied by a projection matrix.  At the first stage there are alternative approaches to constructing a joint sample, while at the second stage there are alternative projection matrices that can be used.  We now discuss each of these stages in detail.

\subsection{Construction of Unreconciled Forecasts} \label{sec:samplingmethod}
The first stage of our procedure, namely to obtain a matrix $\hat{\bm Y}$ is itself broken down into two steps.  In the first step, each level will be modelled independently with details of these models provided in section~\ref{sec:forecastModel}.  Let $\hat{\bm {Z}}_l$ be a $(f_1/f_l)\times N$ matrix defined similarly to $\hat{\bm Y}$. Then, its columns are observations sampled from the joint predictive distribution but only using nodes in the level $l$, i.e. $f\left({\bm {\bm z}_l^{t+h}}|\mathcal{F}^t_l;\hat{\bm{\theta}}\right)$.  A sample from this joint density can be produced by forming multi-step ahead forecasts in the usual recursive fashion and, as a consequence, the dependence within level is preserved.  In the second step, we consider three alternatives for forming a sample $\hat{\bm Y}$ using all $\hat{\bm {Z}}_l$.  Each of these alternatives can be thought of as capturing the dependence between the elements of $\hat{\bm Y}$ in a different way - the appeal of these methods is that they avoid the challenge of modelling for the dependence explicitly.
\subsubsection{Stacked Sample}
The most straightforward way to form $\hat{\bm Y}$ is to simply concatenate the matrices $\hat{\bm {Z}}^{t+h|t}_l$ which we refer to as the `stacked' sample.
\begin{equation}
\hat{\bm Y^{S}}=
\begin{bmatrix}
\hat{\bm Z}_1\\
\hat{\bm Z}_2\\
\vdots\\
\hat{\bm Z}_L
\end{bmatrix}
\end{equation}
Using this approach leads to a joint distribution that preserves the dependence within each level but effectively assumes independence between levels.

\subsubsection{Ranked Sample}
An alternative to the stacked sample involves ordering the elements in each row of $\hat{\bm Y}^S$ in ascending (or descending) order after concatenation.  We refer to this as the `ranked sample' denoted $\hat{\bm Y}^R$.  The rows of $\hat{\bm Y}^R$ will have a comonotonic dependence structure with respect to one another, and this approach can therefore be expected to work well in applications where dependence is high.  Furthermore, the $i^{th}$ column of $\hat{\bm Y}^R$ can be thought of as a vector of the $(i/N)^{th}$ quantiles, each element corresponding to a different node.  As such, this approach also has an interpretation as a method that reconciles quantiles. This approach also has similarities to the combination of probabilistic forecasts by \cite{Lichtendahl2013IsQuantiles}.  Whereas they focus on combining probabilistic forecasts that come from different models, the same idea can easily be applied to appropriately rescaled temporal hierarchies since the probabilistic forecast at each node can be understood as coming from a different model for modelling the wind power over a one hour period.  \cite{Lichtendahl2013IsQuantiles} also propose an approach that averages cumulative probabilities, but find this approach to be inferior to a quantile averaging approach.  Our own application of probability averaging to the reconciliation of temporal hierarchies leads to the same conclusion and these results are omitted.

\subsubsection{Permuted Sample}
A final alternative would be to randomly shuffle the elements within each row  of $\hat{\bm Y}^S$.  We refer to this as the `permuted sample' $\hat{\bm Y}^P$.  The shuffling has the effect of decoupling the dependence within each level, making the rows of $\hat{\bm Y}^P$ independent with respect to one another.  Although this may seem to be an unreasonable approach, it provides an interesting contrast with the ranked sample and may be a useful method that guards against over-fitting when dependence is low.

\subsection{Reconciliation Methods}\label{sec:reconmethods}
Once the matrix $\hat{\bm Y}$ has been formed either as the stacked, ranked or permuted sample, it is pre-multiplied by a projection matrix ${\bm S}{\bm P}$ to yield a reconciled sample.  We consider several alternatives for ${\bm P}$ in this section.
\subsubsection{Bottom Up (BU)}\label{sec:BU}
A simple choice for ${\bm P}$ is to simply ignore information above the bottom level of the hierarchy and simply aggregate the unreconciled bottom level forecasts.  For the example, in Figure \ref{fig:thierarchy_notation} this implies: 
\begin{equation}\label{eq:BUWeight}
\quad
\bm{P}_{BU}=\begin{bmatrix}
    0 & 0 & 0 & 1 & 0 & 0 & 0 \\
    0 & 0 & 0 & 0 & 1 & 0 & 0 \\
    0 & 0 & 0 & 0 & 0 & 1 & 0 \\
    0 & 0 & 0 & 0 & 0 & 0 & 1 \\
  \end{bmatrix},
\end{equation}
or more generally ${\bm P}_{BU}=\begin{bmatrix}{\bm 0}_{m\times (M-m)}&{\bm I}_m\end{bmatrix}$, where ${\bm 0}_{a\times b}$ denotes a $a\times b$ matrix of zeroes and ${\bm I}_{a}$ is an identity matrix of order $a$, 

\subsubsection{Bottom Average (BA)}\label{sec:BA}
Another straightforward method that only uses bottom level information is to average the bottom level.  In this case, the ${\bm P}$ matrix is given by
\begin{equation}\label{eq:LAWeight}
\quad
\bm{P}_{BA}=\begin{bmatrix}
    0 & 0 & 0 & 1/4 & 1/4 & 1/4 & 1/4 \\
    0 & 0 & 0 & 1/4 & 1/4 & 1/4 & 1/4 \\
    0 & 0 & 0 & 1/4 & 1/4 & 1/4 & 1/4 \\
    0 & 0 & 0 & 1/4 & 1/4 & 1/4 & 1/4 \\
  \end{bmatrix},
\end{equation}
in Figure \ref{fig:thierarchy_notation} and by $\bm{P}_{BA}=\begin{bmatrix}{\bm 0}_{m\times (M-m)}&(1/m){\bm 1}_{m\times m}\end{bmatrix}$ in general, where ${\bm 1}_{a\times b}$ denotes a $a\times b$ matrix of ones.

\subsubsection{Global Average (GA)}\label{sec:GA}
Another method is to use information at all nodes of the hierarchy via a simple average, or
\begin{center}
$\tilde{\bm{Y}}_{i,.}=\frac{1}{M}\sum\limits_{j=1}^{M} \hat{\bm{Y}}_{j,.}\quad \forall i$.
\end{center}
This is equivalent to assuming that the matrix ${\bm P}$ is a matrix of ones scaled by $(1/M)$, that is ${\bm P}_{GA}=(1/M){\bm 1}_{m\times M}$. We note that each of the bottom average and global average lead to probabilistic forecasts that are the same for every node, before being transformed back to the original scale.

\subsubsection{Lineal Average (LA)}\label{sec:LA}
An alternative to reconciliation based on an average of all nodes is to build an average based on a set of nodes constructed in the following way.  Supposing we are interested in Node $i$ at the bottom level, take the parent nodes of Node $i$ recursively as well as Node $i$, and calculate the average over the nodes, referring to this as the `lineal average'. The ${\bm P}$ matrix for the lineal average, where Row $i$ allows to take the average of Node $i$ and its ancestor nodes, is defined for the hierarchy in Figure~\ref{fig:thierarchy_notation} as 
\begin{equation}\label{eq:LAWeight}
\quad
\bm{P}_{LA}=\begin{bmatrix}
    1/L & 1/L & 0 & 1/L & 0 & 0 & 0 \\
    1/L & 1/L & 0 & 0 & 1/L & 0 & 0 \\
    1/L & 0 & 1/L & 0 & 0 & 1/L & 0 \\
    1/L & 0 & 1/L & 0 & 0 & 0 & 1/L \\
  \end{bmatrix}.
\end{equation}
This method does not use information from forecasts of sibling nodes to reconcile probabilistic forecasts. A motivation for this is that in the temporal forecasting context, the dependence within each level can be easily preserved. 

\subsubsection{Weighted Least Squares (WLS)}\label{sec:WLS}
In the context of point forecasts, \citet{Athanasopoulos2017ForecastingHierarchies} derive unbiased optimal point forecasts as $\tilde{\bm{Y}}=\bm{S}(\bm{S}'\bm{\Sigma}^{-1}\bm{S})^{-1}\bm{S}'\bm{\Sigma}^{-1}\hat{\bm{Y}}$, where $\bm{\Sigma}$ is the variance covariance matrix of the so-called reconciliation errors. Since $\bm{\Sigma}$ is unidentified \citep{Wickramasuriya2017} it is replaced with a one of three diagonal matrices $\bm{W}$.  Our choice of ${\bm W}$ is similar to the structural scaling approach discussed in \cite{Athanasopoulos2017ForecastingHierarchies}. The only difference between the structural scaling approach and our own is that for the former, the element on the diagonal of ${\bm W}$ corresponding to a node in level $l$ is set to $\bm{f}_l$ while we prefer $\bm{f}_l^2$ reflecting the fact that ${\bm W}$ is a proxy for a variance covariance matrix and that standard deviations rather than variances scale proportionally when the underlying random variable is rescaled.  Furthermore, our choice of ${\bm W}$ leads to results that are equivalent to OLS on the rescaled data while the structural scaling of \cite{Athanasopoulos2017ForecastingHierarchies} does not have this property.

\subsubsection{Cross-Validated (CV)}\label{sec:CV}
A shortcoming of all the approaches above is that the weights are fixed. In this section we propose a class of data-driven weights that are determined via cross-validation to maximise the sharpness of the reconciled predictive distributions, subject to calibration. The notions of sharpness and calibration are discussed by \citet{Gneiting2014ProbabilisticForecasting}. To the best of our knowledge, such a use of cross-validation weights has not been considered in hierarchical reconciliation, either in point forecasting, nor probabilistic forecasting.

The cross-validation procedure involves splitting the sample into three non-overlapping samples, the training sample $\mathcal{T}_{train}$, the validation sample $\mathcal{T}_{val}$ and the test sample $\mathcal{T}_{test}$. Before cross-validation, model parameters are estimated using only training data. We denote these estimates as $\hat{\bm\theta}_{train}$.  Then for all $t+h$ in the validation sample, a sample is produced from $\hat{F}(\bm{y}^{t+h}|\mathcal{F}^t_l;\hat{\bm\theta}_{train})$, where $\hat{F}$ is used to denote the unreconciled predictive cumulative distribution function (CDF). After pre-multiplication by some matrix ${\bm S}{\bm P}$, a sample from the reconciled CDF $\tilde{F}(\bm{y}^{t+h}|\mathcal{F}^t_l;\hat{\bm\theta}_{train})$ is obtained. Let $\tilde{F}_{j,\bm{f}_l}^{t+h}$ be the CDF of the margin corresponding to the $j^{th}$ node in the level $l$ of the hierarchy.  Finally let $R(F,z)$ be a strictly proper scoring rule where $F$ is a predictive CDF, and $z$ is a scaled realisation. 

The objective function for our cross validation is given by
\begin{equation}
CV({\bm P})=L^{-1}\sum\limits_{l=1}^L CV_l({\bm P})\,,
\end{equation}
where
\begin{equation}
CV_l({\bm P})=(\bm{f}_1/\bm{f}_l)^{-1}\sum\limits^{(\bm{f}_1/\bm{f}_l)}_{j=1}\sum\limits_{t+h\in\mathcal{T}_{val}} R(\tilde{F}^{t+h}_{j,\bm{f}_l},z^{t+h}_{j,{\bm{f}_l}}).
\end{equation}\label{eq:CVObj}

In this paper, the scoring function used is the continuous ranked probability score (CRPS) given in general by
\begin{equation}
R(F,z)=\int_u (F(u)-\mathbbm{1}\left\{z\leq u\right\})^2 du\,,
\end{equation}
where $\mathbbm{1}\left\{.\right\}$ is an indicator function equal to 1 if the statement in braces is true and 0 otherwise.  We note that the same scoring rule is used in our empirical evaluation with the notable difference that after determination of cross validation weights, a new $\tilde{F}$ based on both training and validation samples can be obtained.

The quantity $CV({\bm P})$ is optimised with respect to $\bm{P}$. Since the ${\bm P}$ matrix can be quite large we propose the following sparse structure
\begin{equation}\label{eq:CVAll}
\quad
\bm{P}_{CV}=\begin{bmatrix}
    v_{1,1} & v_{2,1} & 0 & v_{3,1} & 0 & 0 & 0 \\
    v_{1,1} & v_{2,1} & 0 & 0 & v_{3,2} & 0 & 0 \\
    v_{1,1} & 0 & v_{2,2} & 0 & 0 & v_{3,3} & 0 \\
    v_{1,1} & 0 & v_{2,2} & 0 & 0 & 0 & v_{3,4} \\
  \end{bmatrix}, 
\end{equation}
where $v_{l,r}$ corresponds to the weight on the $r^{th}$ node in the level $l$. The bottom-up method in Section \ref{sec:BU} is a special case of this method, where only $v_{L,.}$ are $1$, and the other weights are zero. The lineal average method in Section \ref{sec:LA} is also a special case of this method, where all $v$ are $1/L$. 

If the temporal hierarchy of interest is not too large and the study involves a sufficient cross-validation period, all weights of $\bm{P}_{CV}$ could be determined with cross-validation. Where cross-validation is not feasible, further constraints can be placed on the CV weights.  One such restriction is to force the same value of the weight within each level, which gives the following ${\bm P}$ matrix for the hierarchy in Figure~\ref{fig:thierarchy_notation}:

\begin{equation}\label{eq:CVLevel}
\quad
\bm{P}_{CVR}=\begin{bmatrix}
    v_1 & v_2 & 0 & v_3 & 0 & 0 & 0 \\
    v_1 & v_2 & 0 & 0 & v_3 & 0 & 0 \\
    v_1 & 0 & v_2 & 0 & 0 & v_3 & 0 \\
    v_1 & 0 & v_2 & 0 & 0 & 0 & v_3 \\
  \end{bmatrix}, 
\end{equation}
where $v_{l}$ corresponds to the weight on all the nodes in the level $l$. This sparse form reduces the number of weights to optimize over to $L$, with an additional constraint that each row sum is equal to the sum of all the weights. Thus, we use this simpler and practical form of matrix for the case study in Sections \ref{sec:design} and \ref{sec:results}.
To allow for the possibility of poorly calibrated basic forecasts, we tried different restrictions on the weights in cross-validation in Expression \ref{eq:CVLevel}.  In particular we consider the following cases: (1) all weights in a row sum to one and are positive; (2) all weights in a row sum to one; and (3) all weights are unconstrained. 

\begin{figure}[!h]
\centerline{\includegraphics[height=100mm, width=150mm]{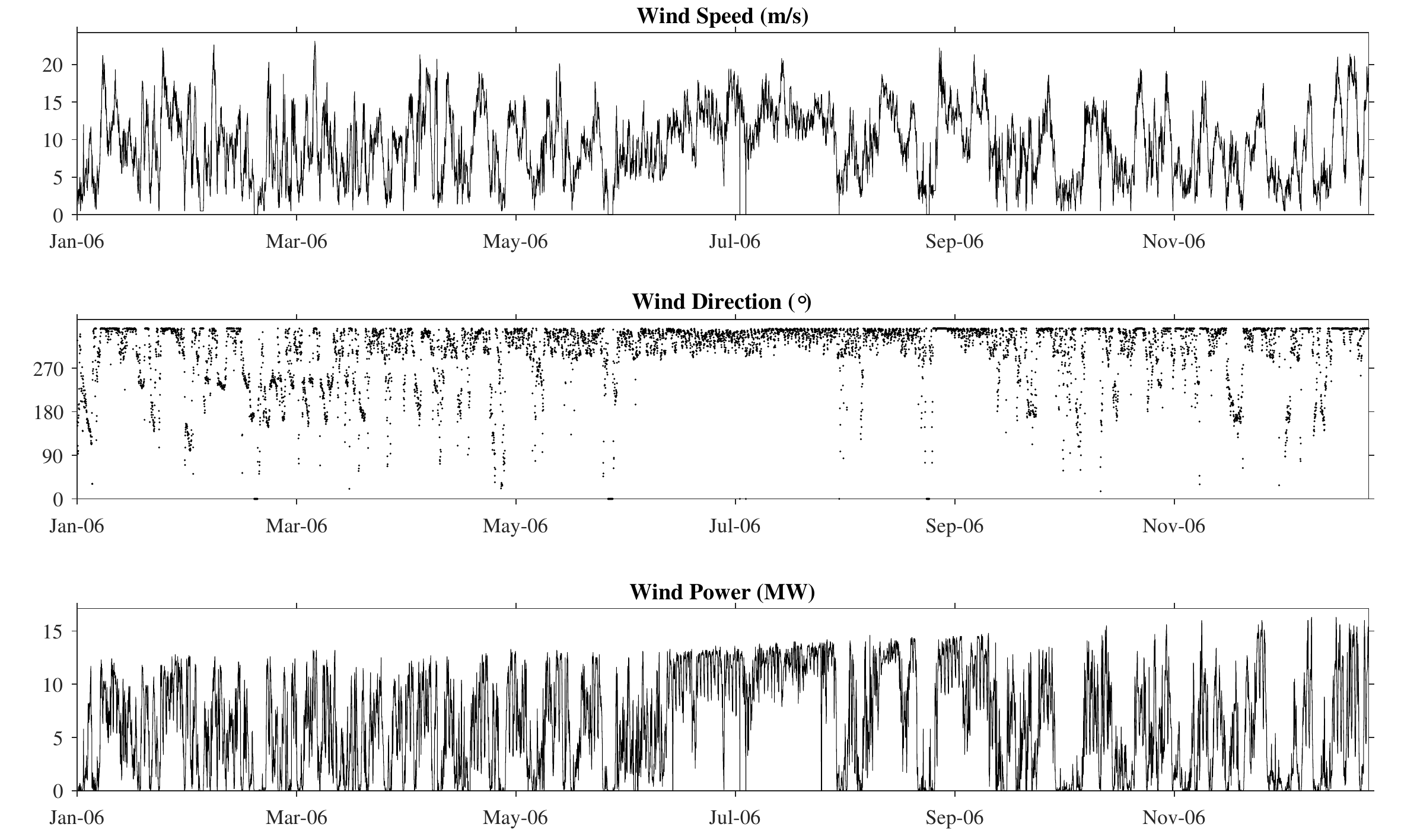}}
\caption{Hourly time series of wind speed, wind direction and wind power in the Rokas wind farm, Crete.}
\label{fig:fig-timeseries-rokas}
\bigskip
\bigskip
\centerline{\includegraphics[width=6in]{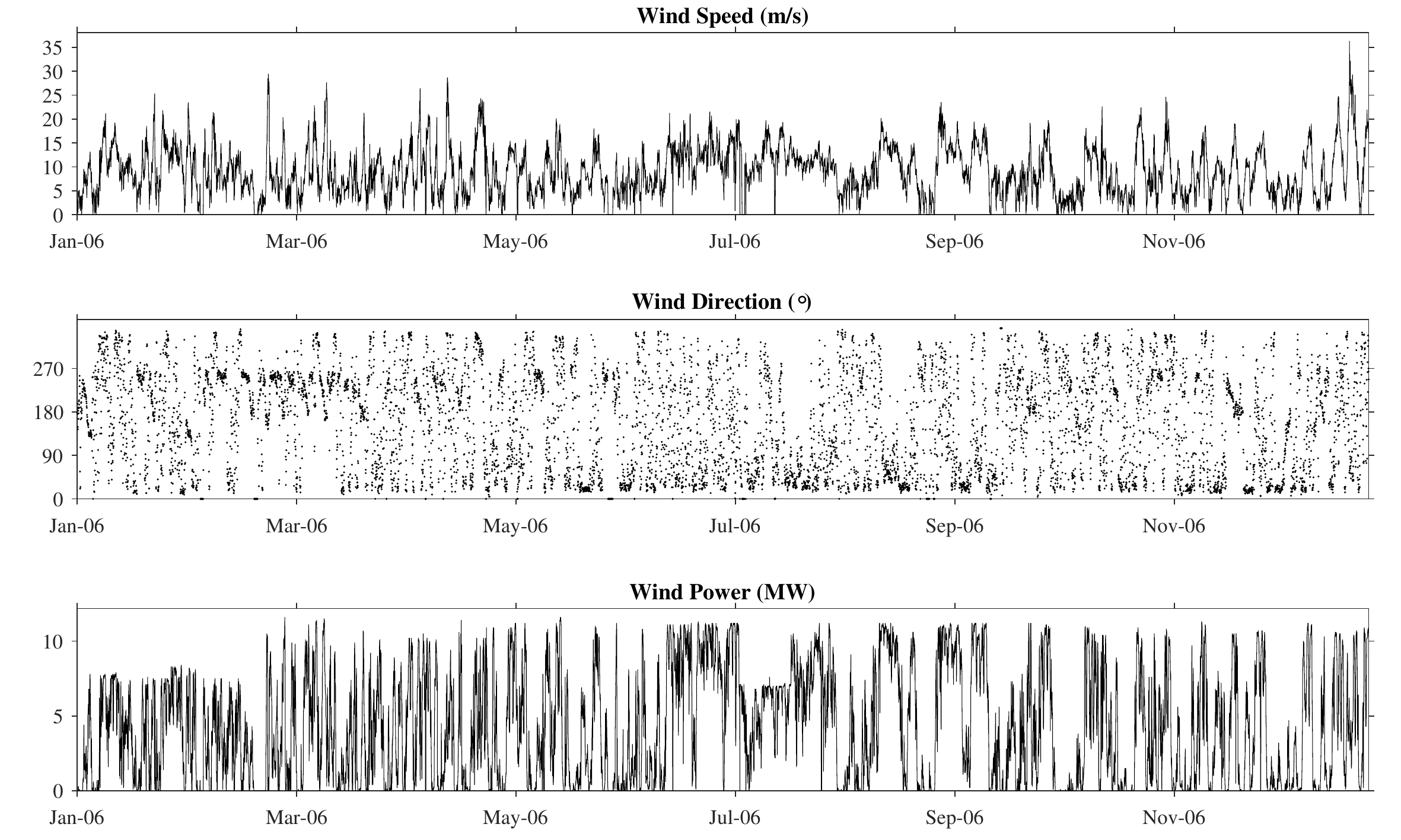}}
\caption{Hourly time series of wind speed, wind direction and wind power in the Aeolos wind farm, Crete.}
\label{fig:fig-timeseries-aeolos}
\end{figure}

\section{Empirical design}\label{sec:design}

\subsection{Temporal Probabilistic Hierarchy of Wind Power}\label{sec:data}
As a case study of the methods we propose in Section \ref{sec:newmethod}, we use hourly time series of wind power from the Rokas and Aeolos wind farms in Crete, the largest island in the Aegean Sea. The island has an autonomous electricity grid and high wind energy potential. The generation capacities of the Rokas and Aeolos wind farms were 16.3MW and 11.6MW, respectively, in 2006. The wind speed and direction observations were recorded at the turbine hub height of the two wind farms and plotted with the corresponding wind power observations in Figures \ref{fig:fig-timeseries-rokas} and \ref{fig:fig-timeseries-aeolos}, respectively, for each hour in 2006, which amounted to 8,760 observations in each series. Each time series was split to $\mathcal{T}_{train}$, the training period of the first 6 months, 1 January 2006 to 30 June 2006, used for training our wind speed density forecasting models; $\mathcal{T}_{val}$, the validation period of the next 3 months, 1 July 2006 to 30 September 2006, used for choosing the most accurate wind speed density forecast model for the time series in each temporal hierarchical level and for selecting the cross-validation weights in Section \ref{sec:CV}; and $\mathcal{T}_{test}$, the test period of the last 3 months, 1 October 2006 to 31 December 2006, reserved for evaluation of the models we proposed. As in Figures \ref{fig:fig-timeseries-rokas} and \ref{fig:fig-timeseries-aeolos}, wind power is more volatile than wind speed, and the volatilities tend to be clustered.

\begin{figure}[!h]
\centerline{\includegraphics[width=6in]{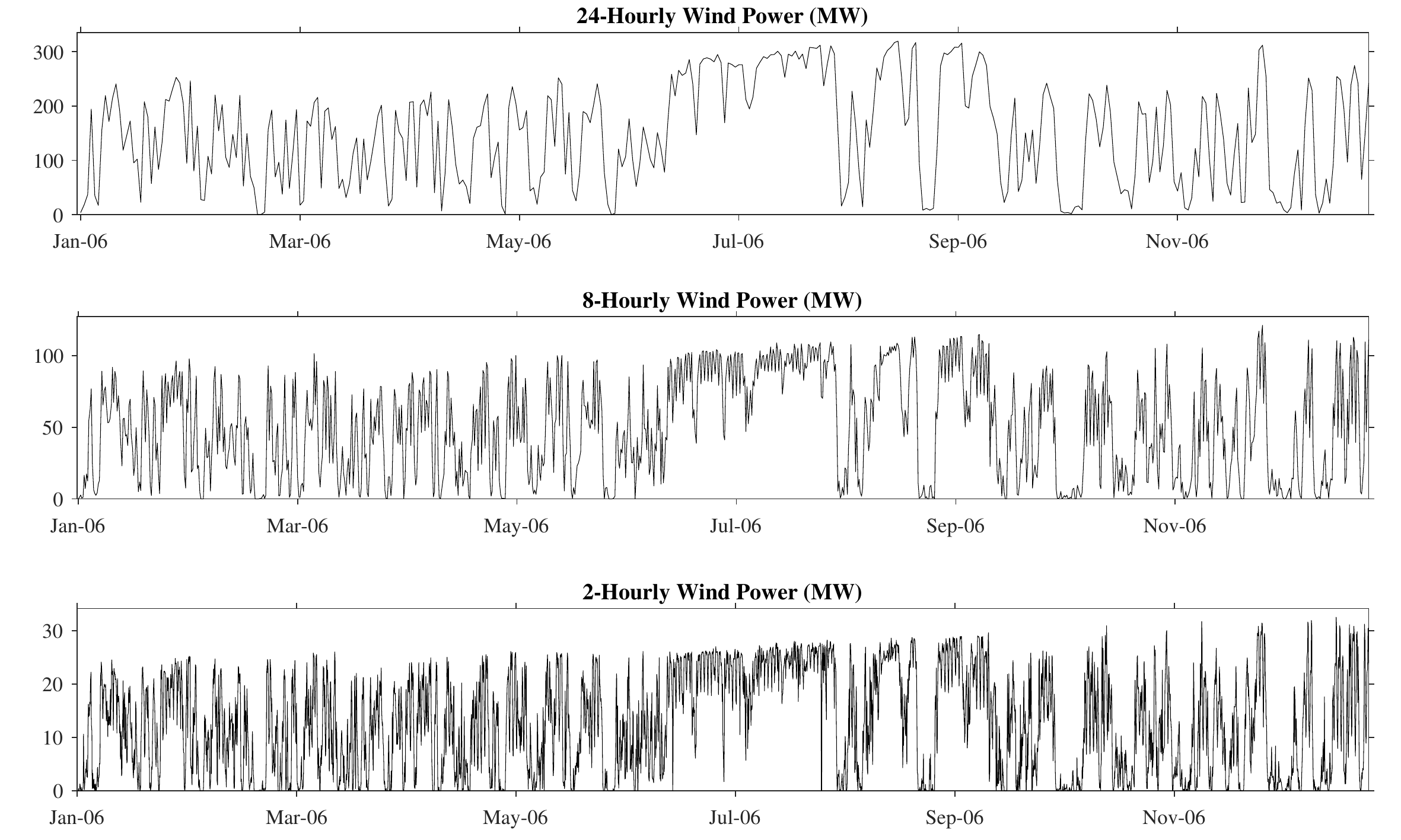}}
\caption{The 24 hourly, 8 hourly and 2 hourly time series of wind power in the Rokas wind farm, Crete.}
\label{fig:fig-timeseries-rokas-multi}
\end{figure}

It is a major challenge for grid operators to maximise the utilisation of wind power due to the intermittency nature of the supply. Due to the inherent uncertainty in the wind power forecasting, probabilistic approaches have received increasing attention recently \citep{Taylor2017ProbabilisticModels,Roulston2003CombiningEnsembles,Gneiting2006CalibratedCenter,Jeon2012UsingForecasting,Hering2010PoweringForecasting,Taylor2015ForecastingEstimation,Dowell2015Very-Short-TermAutoregression}, as these enable more informed decision-making by allowing for the optimal design of bidding strategies and power balance by wind farm operators, grid system operators and electricity traders \citep{Pinson2013WindManagement}. 
One of the most extensive approaches to probabilistic forecasting is to estimate density forecasts, and we estimate these multi-step ahead. Spot power exchange markets are typically a day-ahead auction, and the market price is calculated for each hour of the following day. \citet{Pinson2013WindManagement} also explains that although forecasts up to 2 hours ahead are crucial for dispatch and control problems, much longer lead times are also relevant to decision-making for transmission operations, load-balancing and scheduling for spinning reserve and planning for optimal trading strategies. Therefore, in this paper we focus on enhancing temporal hierarchical probabilistic forecasts up to 24 hours ahead. The overlapping hierarchy consists of 1 $\times$ 24 hour forecast, 2 $\times$ 12 hourly forecasts, 3 $\times$ 8 hourly forecasts, 4 $\times$ 6 hourly forecasts, 6 $\times$ 4 hourly forecasts, 8 $\times$ 3 hourly forecasts, 12 $\times$ 2 hourly forecasts and 24 $\times$ 1 hourly forecasts. This amounts to $L=8$ levels, $M=60$ nodes and $m=24$ bottom-level nodes in the hierarchy. Thus, ${\bm S}$ has the following structure:
\begin{equation}
  {\bm S}=\begin{bmatrix}
  24^{-1} {\bm\iota}'_{24}\\
  12^{-1}{\bm I}_2\otimes  {\bm\iota}'_{12}\\
  8^{-1}{\bm I}_3\otimes  {\bm\iota}'_8\\
  6^{-1}{\bm I}_4\otimes  {\bm\iota}'_6\\
  4^{-1}{\bm I}_6\otimes  {\bm\iota}'_4\\
  3^{-1}{\bm I}_8\otimes  {\bm\iota}'_3\\
  2^{-1}{\bm I}_{12}\otimes  {\bm\iota}'_2\\
  {\bm I}_{24}
  \end{bmatrix}\,,
\end{equation}

\noindent where ${\bm \iota}_{a}$ is an column of $a$ ones and $\otimes$ denotes the Kronecker product. Figure \ref{fig:fig-timeseries-rokas-multi} illustrates the 24 hourly, 8 hourly and 2 hourly time series of Rokas, aggregated from the 1 hourly time series. A lower frequency time series exhibits more smoothed movements.

\subsection{Probabilistic Forecasting Models for Wind Power}\label{sec:forecastModel}To construct a temporal hierarchy of wind power forecasts, we produced density forecasts for each level of the hierarchy. Although a separate density forecast model could be considered for each of the 60 different nodes in the hierarchy, we feel it is unlikely to do so in practice as it demands high computation cost, and significant autocorrelations between nodes in the same level attract modelling them together. In this paper, for each level, a separate multi-step ahead density forecast model is selected after comparing various wind power density forecast models. This procedure is consistent with the study by \cite{Athanasopoulos2017ForecastingHierarchies} that focuses on the point forecasts of temporal hierarchies. 

Statistical models are considerably cheaper than a numerical weather prediction (NWP) system \citep[see, for example,][]{Sloughter2010ProbabilisticAveraging}, and are considered to be very competitive for short lead times \citep{Pinson2013WindManagement}. The statistical models we can consider include direct modelling of wind power using historical simulations, but, as discussed by \citet{Jeon2012UsingForecasting}, it is somewhat challenging due to the non-linear evolution of the time series, and did not perform better than indirect modelling of wind power, which models wind speed first and then converts to wind power. The indirect models we used included univariate autoregressive moving average $-$ generalized autoregressive conditional heteroskedasticity (ARMA$-$GARCH) models for wind speed density forecasting, and VEC$-$type bivariate vector autoregressive moving average $-$ generalised autoregressive conditional heteroskedasticity (VARMA$-$GARCH) models proposed by \citet{Bollerslev1988ACovariances} for wind speed and wind direction density forecasting (see \citet{Jeon2012UsingForecasting} for further details). We further considered long memory dependence in the mean of the wind speed time series captured by the autoregressive fractionally integrated moving average (ARFIMA) model of \citet{Granger1980ANDIFFERENCING} and \citet{Hosking1981FractionalDifferencing} and in the volatility modelled by the fractionally integrated generalized autoregressive conditionally heteroskedastic (FIGARCH) model introduced by \citet{Baillie1996FractionallyHeteroskedasticity}. These ARMA-GARCH type models were fitted to $\mathcal{T}_{train}$ with Gaussian, Student \textit{t} and skew \textit{t} distribution assumptions for the noise term. 

For each hierarchical level of each data set, the wind speed (and wind direction) density forecast model producing the smallest average of the CRPS values evaluated for 1 to 24 hourly ahead wind speed forecasts in $\mathcal{T}_{val}$ was chosen and presented in Table \ref{tab:model}. As evidenced in the paper of \citet{Taylor2009WindModels}, the fractional integration in level and volatility was found to be useful for daily wind speed forecasts. For the time series of higher frequency, the bivariate VARMA-GARCH with Student \textit{t} was chosen the most frequently for both wind farms. Overall, the frequent selection of Student \textit{t} or skew \textit{t} rather than Gaussian distribution indicates the conditional distribution of wind speed follows non-Gaussianity. 

The wind speed and direction density forecasts are then converted to wind power density forecasts using the conditional kernel density estimation, as described by \citet{Jeon2012UsingForecasting} to model the conversion uncertainty in the power curve, which relates wind speed and wind direction to wind power. The noise in the power curve could be brought about by changes in air pressure, temperature, precipitation and wind direction, the complexity of the terrain, different behaviour between speed up and down, turbulence in the turbines, the maintenance of them, and errors in measurement amongst other things.

Based on the models that are individually chosen the best for each level of the hierarchical density forecasting for wind power, 1,000 Monte-Carlo simulated sample paths were generated to construct 1 to 24 hour ahead density forecasts for each node of the hierarchy from each forecast origin in $\mathcal{T}_{test}$. We did not re-estimate density forecasting method parameters for wind power as we rolled the forecast origin forward, because it would be unlikely to be done in practice due to high computational cost, and because the focus of the paper is more about hierarchical reconciliation. 

The density forecasts used for the cross-validated method described in Section~\ref{sec:CV} are produced using a similar approach but in $\mathcal{T}_{val}$. The weights for the cross-validated method with the four different constraints using the $\bm{P}_{CVR}$ matrix discussed in Section \ref{sec:CV} are presented in Tables \ref{tab:CVWeightRokas} and \ref{tab:CVWeightAeolos}. It is interesting to see that the highest weight for each method is mostly on the 1 or 2 hourly hierarchical level for Rokas and 1 hourly hierarchical level for Aeolos. This is sensible as wind power data is fast moving with a high degree of intermittence and the lower level (higher frequency) forecasts contain more useful information.  




\begin{table}[h]\footnotesize
  \centering
  \caption{Models chosen for each wind farm and each hierarchical level. Univariate models produce wind speed density forecasts only. Bivariate models produce density forecasts of wind speed and wind direction.}
  \medskip
	\setlength{\extrarowheight}{2pt}
	\renewcommand\multirowsetup{\centering}
    \begin{tabular}{rcc}
        \hline
			\textbf{Method} & \parbox[c]{0.8cm}{\centering Rokas} & \parbox[c]{0.8cm}{\centering Aeolos} \\
		\hline
24 hourly	&		Univariate ARFIMA-FIGARCH with Gaussian       & Univariate ARMA-FIGARCH with Student \textit{t}			\\       
12 hourly	&		Bivariate VARMA-GARCH with Student \textit{t} & Bivariate VARMA-GARCH with Student \textit{t}				\\
8 hourly	&		Bivariate VARMA-GARCH with Student \textit{t} & Univariate ARMA-FIGARCH with Student \textit{t}			\\
6 hourly	&		Bivariate VARMA-GARCH with Student \textit{t} & Bivariate VARMA-GARCH with Student \textit{t}				\\
4 hourly	&		Bivariate VARMA-GARCH with Student \textit{t} & Bivariate VARMA-GARCH with Student \textit{t}				\\
3 hourly	&		Bivariate VARMA-GARCH with Student \textit{t} & Bivariate VARMA-GARCH with Student \textit{t}				\\
2 hourly	&		Univariate ARMA-GARCH-skew \textit{t}         & Bivariate VARMA-GARCH with Student \textit{t}				\\
1 hourly	&		Univariate ARMA-FIGARCH with skew \textit{t}           & Univariate ARMA-GARCH with skew \textit{t}				\\
		\hline
    \end{tabular}
  \label{tab:model}
\end{table}

\begin{table}[htbp]\footnotesize
  \centering
  \caption{Weights($v$) of the CV method in Section \ref{sec:CV} derived for Rokas, determined by minimising the average of the level-wise average CRPS values in the hierarchy. The sum of \textit{v} is the row sum.}
  \medskip
	\setlength{\extrarowheight}{2pt}
	\renewcommand\multirowsetup{\centering}
    \begin{tabular}{lcccccccccc}
        \hline
			\textbf{Method} & \multicolumn{9}{c}{\textbf{Hierarchical level (in mean)}} \\
			& \parbox[c]{0.8cm}{\centering 24h} & \parbox[c]{0.8cm}{\centering 12h} & \parbox[c]{0.8cm}{\centering 8h} & \parbox[c]{0.8cm}{\centering 6h} & \parbox[c]{0.8cm}{\centering 4h} & \parbox[c]{0.8cm}{\centering 3h} & \parbox[c]{0.8cm}{\centering 2h} & \parbox[c]{0.8cm}{\centering 1h} & \parbox[c]{0.8cm}{\centering Sum} \\
		\hline
  \multicolumn{10}{l}{Permuted Sample} \\  \hspace{6mm} $\sum v_i = 1 \; \& \; \forall v_i \geq 0 $  	  & 0.00  & 0.00  & 0.00  & 0.00  & 0.00  & 0.00  & 1.00  & 0.00  & 1.00  \\
\hspace{6mm} $\sum v_i = 1 $ 	  & -0.37  & 0.05  & 0.38  & -0.15  & 0.19  & 0.10  & 0.87  & -0.07  & 1.00  \\
  \hspace{6mm} Unconstrained 	  & -0.28  & -0.03  & 0.44  & -0.19  & 0.20  & 0.09  & 0.87  & -0.05  & 1.05  \\
 \hdashline \multicolumn{10}{l}{Stacked Sample} \\  \hspace{6mm} $\sum v_i = 1 \; \& \; \forall v_i \geq 0 $ 	  & 0.00  & 0.00  & 0.01  & 0.00  & 0.02  & 0.00  & 0.98  & 0.00  & 1.00  \\
\hspace{6mm} $\sum v_i = 1 $ 	  & -0.34  & 0.29  & 0.23  & -0.07  & 0.49  & -0.38  & 0.64  & 0.14  & 1.00  \\
 \hspace{6mm} Unconstrained 	  & -0.25  & -0.08  & 0.62  & -0.11  & 0.53  & -0.61  & 0.94  & -0.04  & 1.00  \\
 \hdashline \multicolumn{10}{l}{Stacked Sample} \\  \hspace{6mm} $\sum v_i = 1 \; \& \; \forall v_i \geq 0 $ 	  & 0.00  & 0.00  & 0.00  & 0.00  & 0.00  & 0.22  & 0.00  & 0.78  & 1.00  \\
\hspace{6mm} $\sum v_i = 1 $ 	  & -0.01  & -0.04  & 0.04  & -0.02  & 0.03  & 0.07  & 0.38  & 0.56  & 1.00  \\
 \hspace{6mm} Unconstrained 	  & -0.01  & 0.00  & 0.05  & -0.03  & 0.08  & 0.24  & 0.01  & 0.35  & 0.69  \\

   		\hline
    \end{tabular}
  \label{tab:CVWeightRokas}
\end{table}

\begin{table}[htbp]\footnotesize
  \centering
  \caption{Weights($v$) of the CV method in Section \ref{sec:CV} derived for Aeolos, determined by minimising the average of the level-wise average CRPS values in the hierarchy. The sum of \textit{v} is the row sum.}
  \medskip
	\setlength{\extrarowheight}{2pt}
	\renewcommand\multirowsetup{\centering}
    \begin{tabular}{lcccccccccc}
        \hline
			\textbf{Method} & \multicolumn{9}{c}{\textbf{Hierarchical level (in mean)}} \\
			& \parbox[c]{0.8cm}{\centering 24h} & \parbox[c]{0.8cm}{\centering 12h} & \parbox[c]{0.8cm}{\centering 8h} & \parbox[c]{0.8cm}{\centering 6h} & \parbox[c]{0.8cm}{\centering 4h} & \parbox[c]{0.8cm}{\centering 3h} & \parbox[c]{0.8cm}{\centering 2h} & \parbox[c]{0.8cm}{\centering 1h} & \parbox[c]{0.8cm}{\centering Sum} \\
		\hline
 \multicolumn{10}{l}{Permuted Sample} \\  \hspace{6mm} $\sum v_i = 1 \; \& \; \forall v_i \geq 0 $  	  & 0.00  & 0.00  & 0.75  & 0.00  & 0.00  & 0.00  & 0.00  & 0.25  & 1.00  \\
\hspace{6mm} $\sum v_i = 1 $ 	  & -0.01  & -0.00  & 0.53  & -0.34  & -0.00  & 0.34  & -0.19  & 0.68  & 1.00  \\
  \hspace{6mm} Unconstrained 	  & 0.06  & -0.23  & 0.46  & -0.43  & -0.09  & 0.18  & 0.00  & 0.86  & 0.82  \\
 \hdashline \multicolumn{10}{l}{Stacked Sample} \\ \hspace{6mm} $\sum v_i = 1 \; \& \; \forall v_i \geq 0 $ 	  & 0.00  & 0.00  & 0.16  & 0.00  & 0.00  & 0.00  & 0.00  & 0.84  & 1.00  \\
 \hspace{6mm} $\sum v_i = 1 $ 	  & 0.17  & 0.14  & 0.16  & -0.34  & -0.40  & 0.48  & -0.03  & 0.82  & 1.00  \\
 \hspace{6mm} Unconstrained 	  & 0.18  & -0.13  & 0.24  & -0.34  & -0.59  & 0.72  & -0.00  & 0.78  & 0.86  \\
 \hdashline \multicolumn{10}{l}{Stacked Sample} \\ \hspace{6mm} $\sum v_i = 1 \; \& \; \forall v_i \geq 0 $ 	  & 0.00  & 0.00  & 0.01  & 0.00  & 0.00  & 0.00  & 0.00  & 0.99  & 1.00  \\
\hspace{6mm} $\sum v_i = 1 $ 	  & 0.01  & -0.01  & 0.03  & -0.05  & -0.01  & 0.02  & 0.02  & 1.00  & 1.00  \\
  \hspace{6mm} Unconstrained 	  & -0.01  & -0.02  & 0.03  & -0.07  & 0.13  & 0.04  & 0.32  & 0.34  & 0.77  \\

   		\hline
    \end{tabular}
  \label{tab:CVWeightAeolos}
\end{table}

\section{Empirical results}\label{sec:results}

In this section, the methods suggested in Section \ref{sec:newmethod} were compared in terms of the accuracy in the density forecast and the point forecast, for the hierarchy in the case study defined in Section \ref{sec:design}. 

\subsection{Density Forecast Evaluation}\label{sec:densityEvaluation}

For the stacked, ranked and permuted samples and for each reconciliation method, the CRPS value of the wind power density forecast from each of 60 nodes are evaluated for each forecast origin in $\mathcal{T}_{test}$. These values are averaged across $\mathcal{T}_{test}$, and then averaged again across all the nodes in each hierarchical level to be presented as each column of Table \ref{tab:crps}. The final column of the table is the average of all the previous columns in the same row, equivalent to the average of the level-wise average of the CRPS values in the hierarchy. The unit of the CRPS values is Mega Watt (MW), and lower values of this measure are preferred.

If we look at Table \ref{tab:crps}, the CRPS values are presented first by the sampling scheme defined in Section \ref{sec:samplingmethod}, namely the permuted, stacked or ranked sample,  and then by the reconciliation methods, defined in Section \ref{sec:reconmethods}, as the results are more influenced by the choice of the sampling scheme than the choice of the reconciliation method. For example, if we look at the final column, the permuted sample, which would make sense for independent data, and the stacked sampling scheme, which uses the level-wise dependency given by Monte-Carlo simulations of underlying density forecast generation process, mostly did not improve the CRPS of the benchmark, no-reconciliation even though we applied various reconciliation methods. The only exception was the cross-validated reconciliation methods of the permuted sample, which were outperforming no-reconciliation in the levels from 24 hourly to 4 hourly, but not for 2 hourly and 1 hourly. It is particularly disappointing to see the stacked sample demonstrated little improvement over the permuted sample in bottom-up, bottom average, global average, lineal average and WLS, while it was much worse in cross-validated. On the other hand, if we look at the CRPS values in the ranked sample, all the reconciliation methods clearly improved the results of no-reconciliation. The strong performance of the ranked sample may be explained by its interpretation as a forecast combination method.

In the ranked sample, the greatest accuracy was obtained by cross-validated for every level. Cross-validated synthesises information from every level based on 'data-driven' cross-validation weights presented in Tables \ref{tab:CVWeightRokas} and \ref{tab:CVWeightAeolos}, which clearly improved the overall density forecasting performance over the other reconciliation methods. We could not find any consistent difference between various cross-validation conditions in the ranked sample. In terms of accuracy, cross-validated is followed by global average, WLS, bottom average, bottom-up and lineal average. Given that lineal average is a special case of cross-validated, where all $v$ are $1/L$, the poor performance of lineal average in comparison with cross-validated suggests that the optimal weights are far from such fixed weights. It is surprising to see that global average, bottom-up and bottom average performed well, given their simplicity. 

\begin{table}[h]\footnotesize
  \centering
  \caption{CRPS measured for each level of the hierarchy, averaged over the Rokas and Aeolos wind farms.}
	\setlength{\extrarowheight}{2pt}
	\renewcommand\multirowsetup{\centering}
    \begin{tabular}{lcccccccccc}
        \hline
			\multirow{2}{55mm}{\textbf{Sampling Scheme \& Reconciliation Method}} & \multicolumn{9}{c}{\textbf{Forecast Resolution of Each Level}} \\
			& \parbox[c]{0.8cm}{\centering 24h} & \parbox[c]{0.8cm}{\centering 12h} & \parbox[c]{0.8cm}{\centering 8h} & \parbox[c]{0.8cm}{\centering 6h} & \parbox[c]{0.8cm}{\centering 4h} & \parbox[c]{0.8cm}{\centering 3h} & \parbox[c]{0.8cm}{\centering 2h} & \parbox[c]{0.8cm}{\centering 1h} & \parbox[c]{0.8cm}{\centering Mean} \\
		\hline
 No-reconciliation 	  & 1.69  & 1.75  & 1.72  & 1.78  & 1.76  & 1.76  & 1.70  & 1.74  & 1.74  \\
 \hdashline \multicolumn{10}{l}{Permuted Sample} \\ \hspace{3mm}  Bottom-up	  & 1.59  & 1.84  & 1.96  & 1.89  & 1.89  & 1.86  & 1.80  & 1.75  & 1.82  \\
 \hspace{3mm}  Bottom Average 	  & 1.59  & 1.93  & 2.12  & 2.11  & 2.21  & 2.24  & 2.29  & 2.34  & 2.10  \\
 \hspace{3mm}  Global Average 	  & 1.73  & 2.05  & 2.25  & 2.24  & 2.34  & 2.38  & 2.42  & 2.48  & 2.24  \\
 \hspace{3mm}  Lineal Average 	  & 1.77  & 2.01  & 2.18  & 2.14  & 2.20  & 2.22  & 2.24  & 2.27  & 2.13  \\
 \hspace{3mm}  WLS	  & 1.73  & 1.99  & 2.13  & 2.07  & 2.07  & 2.02  & 1.95  & 1.79  & 1.97  \\
 \hspace{3mm}  Cross-validated $\sum v_i = 1 \; \& \; \forall v_i \geq 0 $	  & 1.42  & 1.69  & 1.78  & 1.76  & 1.76  & 1.79  & 1.76  & 1.78  & 1.72  \\
 \hspace{28.5mm}  $\sum v_i = 1 $	  & 1.29  & 1.59  & 1.70  & 1.68  & 1.72  & 1.73  & 1.73  & 1.75  & 1.65  \\
 \hspace{28.5mm}  Unconstrained	  & 1.29  & 1.57  & 1.69  & 1.67  & 1.71  & 1.72  & 1.72  & 1.73  & 1.64  \\
 \hdashline \multicolumn{10}{l}{Ranked Sample} \\ \hspace{3mm}  Bottom-up	  & 1.34  & 1.52  & 1.62  & 1.63  & 1.67  & 1.69  & 1.71  & 1.74  & 1.62  \\
 \hspace{3mm}  Bottom Average 	  & 1.34  & 1.52  & 1.61  & 1.62  & 1.66  & 1.69  & 1.71  & 1.74  & 1.61  \\
 \hspace{3mm}  Global Average 	  & 1.32  & 1.50  & 1.60  & 1.61  & 1.65  & 1.68  & 1.70  & 1.73  & 1.60  \\
 \hspace{3mm}  Lineal Average 	  & 1.38  & 1.56  & 1.67  & 1.67  & 1.72  & 1.74  & 1.77  & 1.80  & 1.66  \\
 \hspace{3mm}  WLS	  & 1.32  & 1.50  & 1.61  & 1.61  & 1.65  & 1.68  & 1.70  & 1.73  & 1.60  \\
 \hspace{3mm}  Cross-validated   $\sum v_i = 1 \; \& \; \forall v_i \geq 0 $	  & \textbf{1.27}  & \textbf{1.48}  & \textbf{1.59}  & \textbf{1.59}  & \textbf{1.64}  & \textbf{1.67}  & \textbf{1.69}  & \textbf{1.72}  & \textbf{1.58}  \\
  \hspace{28.5mm} $\sum v_i = 1 $	  & 1.28  & 1.49  & \textbf{1.59}  & \textbf{1.59}  & \textbf{1.64}  & \textbf{1.67}  & \textbf{1.69}  & \textbf{1.72}  & \textbf{1.58}  \\

 \hspace{28.5mm}  Unconstrained	  & 1.28  & 1.49  & 1.60  & \textbf{1.59}  & 1.65  & \textbf{1.67}  & 1.70  & 1.73  & 1.59  \\
 \hdashline \multicolumn{10}{l}{Stacked Sample} \\ \hspace{3mm}  Bottom-up	  & 1.58  & 1.84  & 1.96  & 1.89  & 1.89  & 1.85  & 1.80  & 1.74  & 1.82  \\
 \hspace{3mm}  Bottom Average 	  & 1.58  & 1.93  & 2.12  & 2.11  & 2.21  & 2.24  & 2.29  & 2.34  & 2.10  \\
 \hspace{3mm}  Global Average 	  & 1.73  & 2.05  & 2.25  & 2.24  & 2.34  & 2.37  & 2.42  & 2.48  & 2.23  \\
 \hspace{3mm}  Lineal Average 	  & 1.77  & 2.01  & 2.18  & 2.14  & 2.20  & 2.22  & 2.24  & 2.26  & 2.13  \\
 \hspace{3mm}  WLS	  & 1.73  & 1.99  & 2.13  & 2.07  & 2.07  & 2.02  & 1.95  & 1.79  & 1.97  \\
 \hspace{3mm}  Cross-validated $\sum v_i = 1 \; \& \; \forall v_i \geq 0 $	  & 1.62  & 1.89  & 2.02  & 1.94  & 1.95  & 1.91  & 1.85  & 1.77  & 1.87  \\
 \hspace{28.5mm}  $\sum v_i = 1 $	  & 1.57  & 1.86  & 2.00  & 1.93  & 1.94  & 1.91  & 1.85  & 1.78  & 1.86  \\
 \hspace{28.5mm}  Unconstrained	  & 1.98  & 2.13  & 2.22  & 2.19  & 2.19  & 2.18  & 2.15  & 2.11  & 2.14  \\

		\hline
    \end{tabular}
    \raggedright{Note: Lower values are better. The best value in each column is in bold.}
    \label{tab:crps}
\end{table}

\begin{figure}[h]
    \centerline{\includegraphics[width=7in]{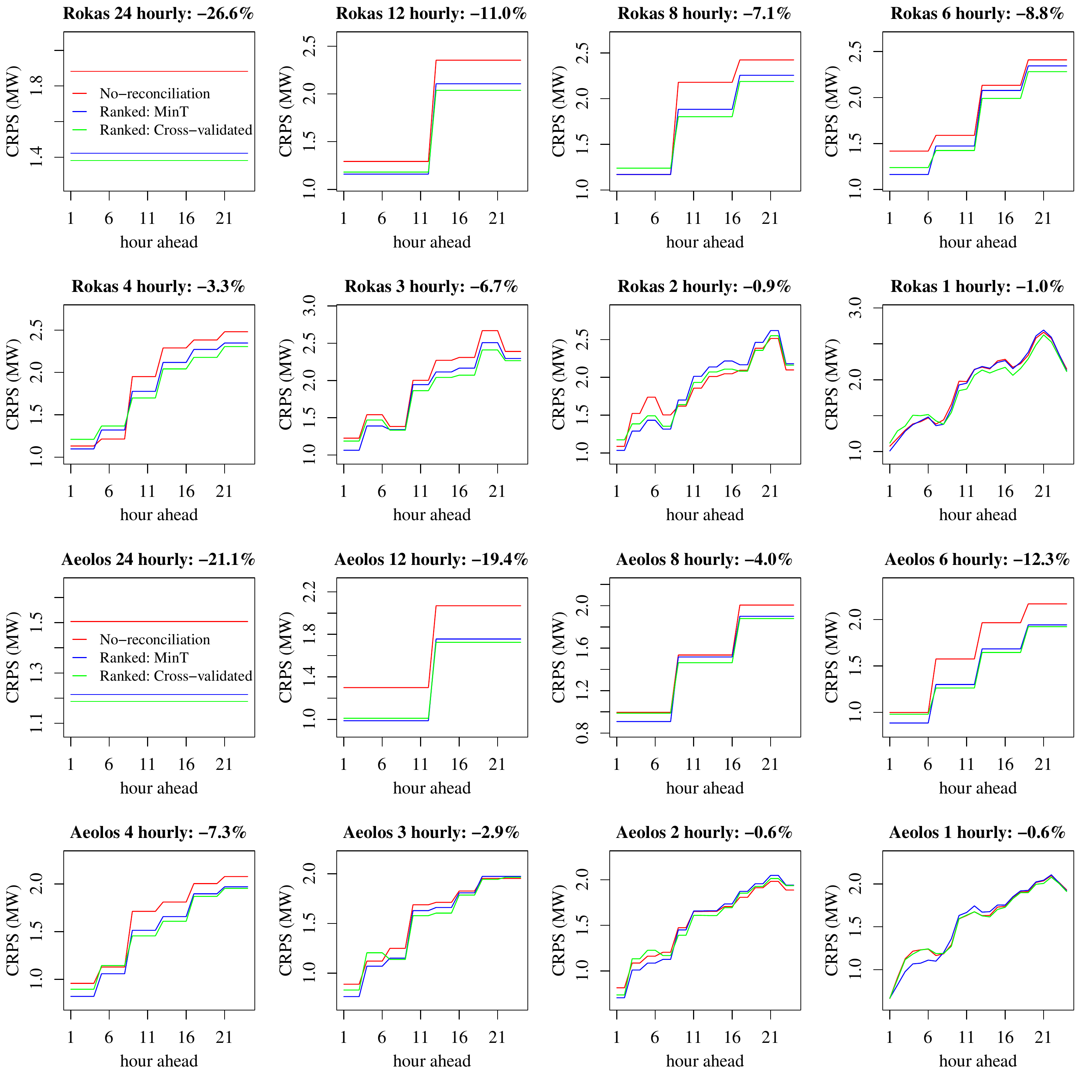}}
\caption{Probabilistic Evaluation of wind power forecasts in the evaluation period using CRPS for the Rokas and Aeolos wind farms, comparing (1) no-reconciliation and (2) bottom-up using the ranked sample and (3)  Cross-validated using the ranked sample with constraint, $\sum v_i = 1 \; \& \; \forall v_i \geq 0 $. The improvement of cross-validated over no-reconciliation is presented in average percentage for each level separately, on top of each plot. Lower values are better.}
\label{fig:fig-crps-both}
\end{figure}

To investigate more closely the performance for each of the forecast lead times in each level and for each wind farm, we plotted in Figure \ref{fig:fig-crps-both} the CRPS values of no-reconciliation, WLS using ranked sample and cross-validated using ranked sample with no-constraint. Although the three months in the evaluation period is not sufficient to obtain smooth lines of CRPS in the plots, there is a clear tendency for the CRPS values to increase with forecast lead times in each plot. The title of each plot in Figure \ref{fig:fig-crps-both} indicates the average improvement of cross-validated over no-reconciliation, in terms of CRPS, where lower values are preferred. For example, the 24 hourly density forecast of cross-validated produced the CRPS values that are 26.6\% and 21.1\% smaller than no-reconciliation in the Rokas and Aeolos wind farms, respectively. As we increase the forecast resolution by moving further down the hierarchical level in the following plots, this enhancement tended to be reduced. This indicates that wind power density forecasts of the higher resolution could be further enhanced by synthesizing forecasts of lower resolution. The reconciliation in some sense `hedges' the misspecification errors by synthesizing information from all hierarchical nodes.

\subsection{Point Forecast Evaluation}\label{sec:pointEvaluation}
Although density forecast performance is our primary concern, a density forecast could
produce a point forecast by calculating the expected value from the density, and this could
be useful for evaluating the centre of the density forecast. \citet{Gneiting2011bMakingEvaluatingPointForecasts,Gneiting2011aQuantilesAsOptimalPointForecasts} explains the median of a density forecast being the optimal point forecast  for symmetric piecewise linear loss functions such as MAE. Indeed, in terms of MAE, using the median showed slightly higher accuracy than the mean in our empirical results. The MAE result using the median is presented in Table \ref{tab:mae}, produced in the same fashion as Table \ref{tab:crps}. 

\begin{figure}[h]
\centerline{\includegraphics[width=7in]{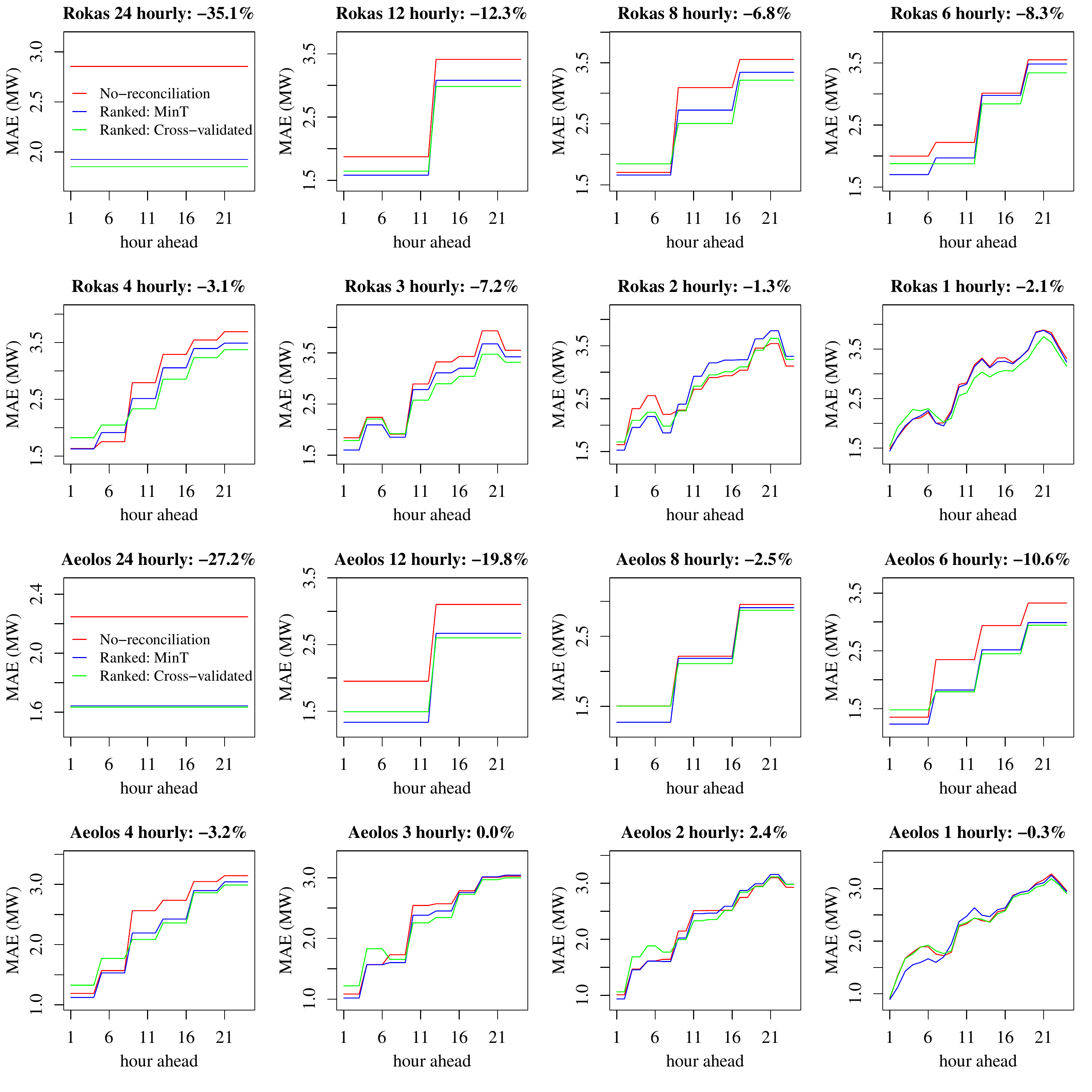}}
\caption{Point Evaluation of wind power density forecasts in the evaluation period using MAE for the Rokas and Aeolos wind farms, comparing (1) no-reconciliation and (2) bottom-up using the ranked sample and (3) Cross-validated using the ranked sample with constraint, $\sum v_i = 1 \; \& \; \forall v_i \geq 0 $. The improvement of cross-validated over no-reconciliation is presented in average percentage for each level separately, on top of each plot. Lower values are better.}
\label{fig:fig-mae-both}
\end{figure}

In Table \ref{tab:mae}, it is surprising to see that the various reconciliation methods we proposed provide more competitive results for MAE than for CRPS. For example, if we look at the overall mean of MAEs in the last column of Table \ref{tab:mae}, most of the combinations of sampling schemes and reconciliation methods produced smaller MAEs than no-reconciliation. This enhancement is clearer in Figure \ref{fig:fig-mae-both}, which is plotted in similar fashion to Figure \ref{fig:fig-crps-both} but using MAE. In the plots for 24 hourly for Rokas and Aeolos, the enhancements of the best density forecast method against no-reconciliation, in terms of MAE, were 35.1\% for Rokas and 27.2\% for Aeolos, whereas the enhancements were 26.6\% and 21.1\% respectively in terms of CRPS. 
This supports the temporal hierarchical density reconciliation methods we propose produce further enhancement in the centre of the forecast distributions. If we go back to Table \ref{tab:mae}, we can observe that the sampling scheme produced the most accurate point forecasts overall was the ranked sample, which was consistent with the results of the density forecast evaluation. Among the ranked sample, the global average and unconstrained cross-validated reconciliation methods were the most accurate.

\begin{table}[h]\footnotesize
  \centering
  \caption{MAE measured for each level of the hierarchy, averaged over the Rokas and Aeolos wind farms.}
	\setlength{\extrarowheight}{2pt}
	\renewcommand\multirowsetup{\centering}
    \begin{tabular}{lcccccccccc}
        \hline
			\multirow{2}{55mm}{\textbf{Sampling Scheme \& Reconciliation Method}} & \multicolumn{9}{c}{\textbf{Forecast Resolution of Each Level}} \\
			& \parbox[c]{0.8cm}{\centering 24h} & \parbox[c]{0.8cm}{\centering 12h} & \parbox[c]{0.8cm}{\centering 8h} & \parbox[c]{0.8cm}{\centering 6h} & \parbox[c]{0.8cm}{\centering 4h} & \parbox[c]{0.8cm}{\centering 3h} & \parbox[c]{0.8cm}{\centering 2h} & \parbox[c]{0.8cm}{\centering 1h} & \parbox[c]{0.8cm}{\centering Mean} \\
		\hline
 No-reconciliation 	  & 2.55  & 2.58  & 2.50  & 2.59  & 2.58  & 2.59  & 2.49  & 2.59  & 2.56  \\
 \hdashline \multicolumn{10}{l}{Permuted Sample} \\ \hspace{3mm}  Bottom-up	  & 1.91  & 2.29  & 2.47  & 2.46  & 2.55  & 2.58  & 2.59  & 2.59  & 2.43  \\
 \hspace{3mm}  Bottom Average 	  & 1.91  & 2.27  & 2.46  & 2.45  & 2.55  & 2.59  & 2.64  & 2.70  & 2.45  \\
 \hspace{3mm}  Global Average 	  & 1.95  & 2.28  & 2.48  & 2.48  & 2.57  & 2.61  & 2.66  & 2.72  & 2.47  \\
 \hspace{3mm}  Lineal Average 	  & 2.08  & 2.37  & 2.58  & 2.58  & 2.67  & 2.70  & 2.75  & 2.80  & 2.57  \\
 \hspace{3mm}  WLS	  & 1.95  & 2.30  & 2.51  & 2.50  & 2.58  & 2.61  & 2.64  & 2.63  & 2.46  \\
 \hspace{3mm}  Cross-validated $\sum v_i = 1 \; \& \; \forall v_i \geq 0 $	  & 1.82  & 2.25  & 2.43  & 2.40  & 2.49  & 2.51  & 2.53  & 2.57  & 2.38  \\
 \hspace{28.5mm}  $\sum v_i = 1 $	  & 1.74  & 2.21  & 2.39  & 2.37  & 2.47  & 2.49  & 2.52  & 2.55  & 2.34  \\
 \hspace{28.5mm}  Unconstrained	  & 1.74  & 2.20  & 2.37  & 2.36  & 2.45  & 2.47  & 2.50  & 2.54  & 2.33  \\
 \hdashline \multicolumn{10}{l}{Ranked Sample} \\ \hspace{3mm}  Bottom-up	  & 1.84  & 2.22  & 2.40  & 2.39  & 2.47  & 2.51  & 2.55  & 2.59  & 2.37  \\
 \hspace{3mm}  Bottom Average 	  & 1.84  & 2.15  & 2.31  & 2.31  & 2.40  & 2.43  & 2.48  & 2.52  & 2.31  \\
 \hspace{3mm}  Global Average 	  & 1.78  & \textbf{2.11}  & \textbf{2.29}  & \textbf{2.28}  & \textbf{2.37}  & \textbf{2.42}  & \textbf{2.46}  & 2.51  & \textbf{2.28}  \\
 \hspace{3mm}  Lineal Average 	  & 1.96  & 2.26  & 2.45  & 2.43  & 2.52  & 2.56  & 2.60  & 2.65  & 2.43  \\
 \hspace{3mm}  WLS	  & 1.78  & 2.17  & 2.35  & 2.34  & 2.43  & 2.47  & 2.51  & 2.56  & 2.33  \\
 \hspace{3mm}  Cross-validated $\sum v_i = 1 \; \& \; \forall v_i \geq 0 $	  & \textbf{1.71}  & 2.19  & 2.35  & 2.33  & 2.42  & 2.45  & 2.50  & 2.54  & 2.31  \\
 \hspace{28.5mm}  $\sum v_i = 1 $	  & 1.74  & 2.18  & 2.34  & 2.33  & 2.42  & 2.45  & 2.49  & 2.53  & 2.31  \\
 \hspace{28.5mm}  Unconstrained	  & 1.76  & 2.16  & 2.31  & 2.29  & 2.39  & \textbf{2.42}  & \textbf{2.46}  & \textbf{2.50}  & 2.29  \\
 \hdashline \multicolumn{10}{l}{Stacked Sample} \\ \hspace{3mm}  Bottom-up	  & 1.91  & 2.29  & 2.47  & 2.46  & 2.55  & 2.57  & 2.58  & 2.59  & 2.43  \\
 \hspace{3mm}  Bottom Average 	  & 1.91  & 2.27  & 2.46  & 2.45  & 2.55  & 2.59  & 2.64  & 2.70  & 2.45  \\
 \hspace{3mm}  Global Average 	  & 1.95  & 2.28  & 2.48  & 2.48  & 2.57  & 2.61  & 2.67  & 2.72  & 2.47  \\
 \hspace{3mm}  Lineal Average 	  & 2.08  & 2.37  & 2.58  & 2.57  & 2.67  & 2.70  & 2.74  & 2.80  & 2.56  \\
 \hspace{3mm}  WLS	  & 1.95  & 2.30  & 2.51  & 2.49  & 2.58  & 2.61  & 2.64  & 2.63  & 2.46  \\
 \hspace{3mm}  Cross-validated $\sum v_i = 1 \; \& \; \forall v_i \geq 0 $	  & 1.92  & 2.31  & 2.50  & 2.48  & 2.57  & 2.60  & 2.61  & 2.61  & 2.45  \\
 \hspace{28.5mm}  $\sum v_i = 1 $	  & 1.86  & 2.27  & 2.46  & 2.44  & 2.54  & 2.56  & 2.58  & 2.60  & 2.41  \\
 \hspace{28.5mm}  Unconstrained	  & 2.20  & 2.43  & 2.57  & 2.59  & 2.64  & 2.68  & 2.71  & 2.74  & 2.57  \\

\hline
    \end{tabular}
    \raggedright{Note: Point forecasts are medians of density forecasts. Lower values are better. The best value in each column is in bold.}
   \label{tab:mae}
\end{table}

\section{Concluding Comments}\label{sec:conclusions}

This paper focused on the reconciliation of probabilistic forecasts that are arranged in hierarchical structures, with a particular focus on temporal hierarchies. We propose three schemes for obtaining samples from the estimates of the joint densities, namely permuted, ranked and stacked sampling. These approaches correspond to the cases of no dependence between the hierarchical nodes, comonotonic dependence between nodes and temporal model driven dependencies within a level respectively. These sampling schemes are then applied to several reconciliation approaches, bottom-up, bottom/global/lineal average and WLS. Furthermore, we investigated for the first time the use of a cross-validation approach for obtaining the reconciliation weights. The performance of the various combinations of sampling schemes and reconciliation methods was subsequently measured by producing and evaluating probabilistic wind power forecasts reconciled from various frequencies. 

The empirical results from two wind farms in Greece suggest that cross-validation reconciliation based on ranked samples offers the best performance compared to all other approaches. Performance enhancement is up to 25\% and up to 35\% relatively to no-reconciliation for density and point forecast evaluation respectively. Lower resolutions (higher levels of aggregation) enjoyed the most performance benefits, providing direct managerial benefits for transmission operations and planning for optimal trading strategies. The results also show that comonotonic aggregation of quantiles worked better than modelling level-wise dependencies.

While our study focused on the application of the various approaches in temporal hierarchies, these can be applied equally to the case of cross-sectional hierarchies, thus extending the work by \cite{Taieb2017CoherentSeries} who investigate the construction of coherent probabilistic forecasts in a bottom-up fashion. Furthermore, our study investigates for the very first time the performance of cross-validated derived weights for the construction of coherent forecasts. We suggest that cross-validation can also be applied to the construction of coherent point forecasts. 

Looking forward, our research also poses new research questions that lie outside the scope of the current paper. For example, although an advantage of the stacked sample, ranked sample and permuted sample is their ease of construction, it may be worthwhile developing more complicated merging schemes based on the dependence structure of in sample forecast errors and investigating whether such schemes lead to better reconciled probabilistic forecasts. It may also be worthwhile investigating whether the sparse structure of the ${\bm P}$ matrix can be selected in a more data driven way, especially for cross sectional hierarchies where a different pattern of sparsity may be required to compensate for the base level forecasts are produced at each node rather than at each level. Finally, it would be interesting to see if methods based on ensemble or physics that can generate density forecasts also produce benefits using (temporal) hierarchical reconciliation.

\section*{Acknowledgements}\label{sec:acknowledgements}

Jooyoung Jeon was supported by the EPSRC grant (EP/N03466X/1). We are grateful to George Sideratos of the National Technical University of Athens and the EU SafeWind Project for providing the data. We are also grateful for the insight comments of participants at the International Symposium on Energy Analytics in Cairns, Australia, 2017.

\clearpage

\appendix


\singlespacing

\bibliographystyle{tPRS}
\bibliography{ref}

\end{document}